\documentstyle[prl,epsf,floats,aps]{revtex}

\newlength{\figwidth}
\def\elab#1#2{\begin{equation}\label{#1}#2\end{equation}}
\def\ealab#1#2{\begin{eqnarray}\label{#1}#2\end{eqnarray}}

\def\eref#1{Eq.~(\ref{#1})}

\def\figref#1{Fig.~(\ref{#1})}
\def\figrefab#1#2{Fig.~(\ref{#1}#2)}

\def\mysection#1{\paragraph*{#1}}

\def\anti#1{\overline#1}

\def\CI#1{\cite{r:#1}}
\def\BI#1{\bibitem{r:#1}}

\def\PRep#1#2#3#4{#1, {\it Phys. Rep.} {\bf#2}, #3 (19#4)}
\def\JPG#1#2#3#4{#1, {\it Jour. Phys.} {\bf G#2}, #3 (19#4)}
\def\APP#1#2#3#4{#1, {\it Astropart. Phys.} {\bf#2}, #3 (19#4)}
\def\PRD#1#2#3#4{#1, {\it Phys. Rev.} {\bf D#2}, #3 (19#4)}
\def\NPB#1#2#3#4{#1, {\it Nucl. Phys.} {\bf B#2}, #3 (19#4)}
\def\PR#1#2#3#4{#1, {\it Phys. Rev.} {\bf #2}, #3 (19#4)}
\def\PRL#1#2#3#4{#1, {\it Phys. Rev. Lett.} {\bf#2}, #3 (19#4)}
\def\ARNPS#1#2#3#4{#1, {\it Ann. Rev. Nucl. Part. Sci.} {\bf #2}, #3 (19#4)}
\def\PLB#1#2#3#4{#1, {\it Phys. Lett.} {\bf #2B}, #3 (19#4)}
\def\APJ#1#2#3#4{#1, {\it Astrophys. J.} {\bf#2}, #3 (19#4)}

\begin{document}

\twocolumn[\hsize\textwidth\columnwidth\hsize\csname
@twocolumnfalse\endcsname

\draft
\title{Neutrino-photon reactions in astrophysics and cosmology}
\author{D. Seckel}
\address{Bartol Research Institute, University of Delaware, 
       Newark DE  19716 \\
Bartol Preprint No. BA-97-32,
Submitted to Phys. Rev. Lett.}
\date{Sept 9, 1997}
\maketitle
%
\begin{abstract}
At energies above the threshold for $W$ production the process $\nu 
\gamma \rightarrow l W^+$ is competive with $\nu \nu$ scattering at the 
same center of mass energies. In a cosmological setting, absorption of 
ultra high energy neutrinos by the microwave photon background is 
comparable to absorption by the neutrino background. In passing through 
matter, the process $\nu \rightarrow l W^+$ will occur in the coulomb 
field of nuclei. For iron, the interaction rate per nucleon is roughly 
20\% of the charge current cross-section. The related process, 
$\anti\nu_e e^- \rightarrow \gamma W^-$ dominates $\anti\nu_e e^-$ 
scattering for about a decade in energy above the resonance for $W$ 
production.
\end{abstract}

\vskip2pc]

%
\narrowtext
%
Neutrinos of very high energy have become a subject of some 
interest\CI{GaisserHS}. Detection of such neutrinos could provide a means 
of identifying and studying sources of the highest energy cosmic 
rays\cite{r:SteckerDSS,r:SzaboP}. Neutrinos, $\gamma$-rays, and nucleons 
are all produced at the source via hadronic reactions. Unlike photons or 
nucleons, however, neutrinos can both escape the central accelerator and 
propagate cosmological distances while preserving line of sight 
information to indicate the location of the source. High energy neutrinos 
from the decay or annihilation of particle dark matter\CI{dark} or from 
the decay of cosmic strings\cite{r:strings,r:horizontal} could provide 
important clues for a deeper understanding of particle physics and/or 
cosmology. Recent models of $\gamma$-ray bursts may be testable by looking 
for coincident neutrinos with energies $E_\nu > 10^{14}$ eV\CI{WaxmanB}.

With these thoughts in mind, experimental efforts are being initiated to detect 
cosmic neutrinos at energies from $10^{14}$ to 
$10^{20}$eV either underwater\cite{r:BAIKAL,r:NESTOR},
underice\cite{r:AMANDA,r:RICE}, or possibly in horizontal 
air showers at an extensive air shower array\CI{horizontal}.
Detection of such high energy neutrinos could be a boon for particle 
physicists as well. It is expected that such neutrinos would be absorbed 
by the Earth. By measuring the flux as a function of nadir angle one 
could measure neutrino-nucleon cross-sections at high 
energies\CI{GandhiQRS}. Such a measurement would supply information about nucleon
structure functions at energies inaccessible to current accelerators. 

In all these cases, estimates of neutrino reaction rates have been based 
upon the exchange of weak vector bosons with nucleons or electrons, or in 
the case of cosmological 
absorption\cite{r:Berezinsky,r:Roulet,r:GondoloGS,r:Yoshida}, 
the cosmic neutrino background. Here 
it is pointed out that neutrino-photon reactions that produce final state 
``on shell" weak vector bosons should not be neglected. The photon can be 
real, as in $\nu \gamma \rightarrow l W^+$, or virtual as in $\nu N 
\rightarrow N l W^+$ catalyzed by the coulomb field of the nucleus. For 
the case of scattering from electrons, the photon may be in the final 
state, $\anti\nu_e e^- \rightarrow \gamma W^-$, which enhances 
$\anti\nu_e e^-$ scattering above the ``Glashow resonance" for $W$ 
production\CI{Glashow}. These three cases are presented below followed by 
a summary and discussion.

\begin{figure}[ht]
\centering\leavevmode
\epsfxsize=\figwidth
\epsfbox{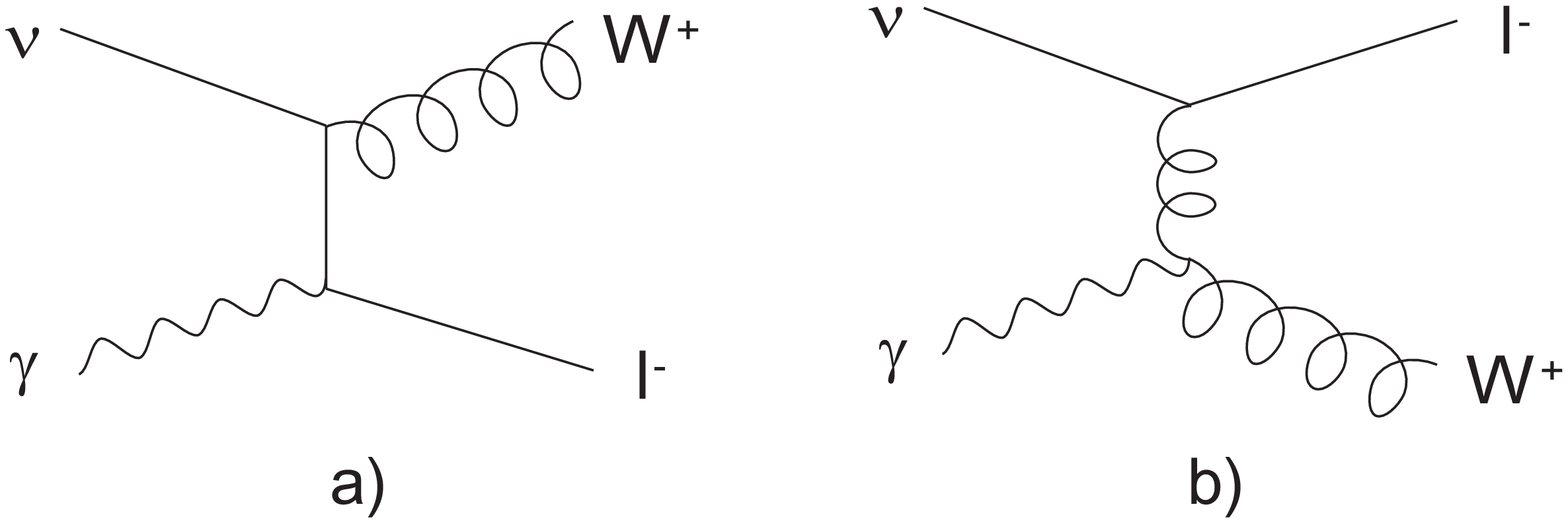}
\smallskip
\caption{Two amplitudes contributing to $\nu \gamma\rightarrow l W^+$:
a) is `Compton like', while b) involves a three gauge coupling. Both must 
be included to maintain gauge invariance.}
\label{fig:fd}
\end{figure}

\mysection{$\nu \gamma \rightarrow \lowercase{l} W^+$.}
It is straightforward to calculate the cross-section for
$\nu \gamma \rightarrow l W^+$ using 
the standard model lagrangian. A general form for the cross-section
is
\elab{eq:sig1}{
d\sigma_{\nu \gamma \rightarrow l W^+} = 
|{\cal M}|^2  \delta^4(p_i - p_f) (2\pi)^4 \frac{1}{4 I} d\rho_f \,,
}
where $p_i$ and $p_f$ denote initial and final particle four-momenta, 
$\rho_f$ is the final particle phase space, $I$ is the lorentz invariant flux 
factor and $\cal{M}$ is the lorentz invariant amplitude for the process.
\figref{fig:fd} shows the two
diagrams that contribute to $\cal{M}$. For the present 
purposes it is sufficient to consider the cross-section for unpolarized 
particles, so $|{\cal M}|^2$ may be simplified by summing over $W$ 
polarizations, $\sum_{\lambda_W} \epsilon^\mu_{\lambda_W}  \epsilon^\nu_{\lambda_W} 
= -(g^{\mu\nu}-\frac{p_W^\mu p_W^\nu}{M_W^2})$. Before performing a 
similar sum over photon polarizations it is useful to calculate the 
electromagnetic current tensor ${\cal J}_{\mu\nu}=J_\mu J_\nu$, where
$J$ is the current which couples to photons. The matrix 
element can then be written in the form $|{\cal M}|^2 = \epsilon^\mu 
\epsilon^\nu {\cal J}_{\mu\nu}$, 
where $\epsilon$ is the photon polarization vector.
For unpolarized photons one then uses 
the average $\frac{1}{2} \sum_{\lambda} \epsilon^\mu_{\lambda} \epsilon^\nu_{\lambda} 
= -\frac{1}{2} g^{\mu\nu}$. As a check of the algebra one can test that 
$p_\gamma^\mu {\cal J}_{\mu\nu} = p_\gamma^\nu {\cal J}_{\mu\nu} = 0$ 
which is demanded by gauge invariance. ${\cal J}$ is
also useful for calculating $\nu N \rightarrow N l W^+$ in the nuclear coulomb 
field.

\begin{figure}[ht]
\centering\leavevmode
\epsfxsize=\figwidth
\epsfbox{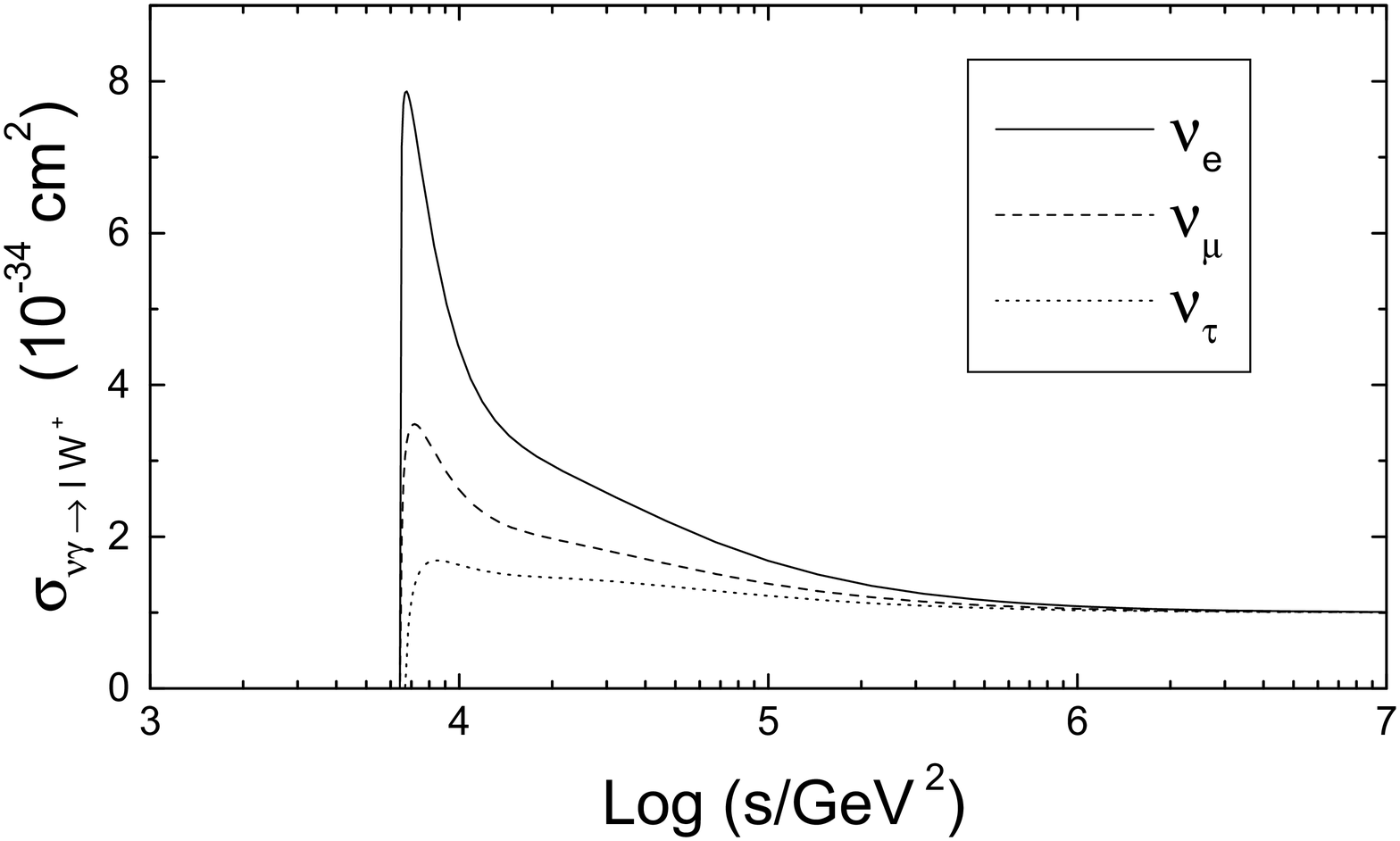}
\smallskip
\caption{Cross-section for $\nu \gamma \rightarrow l W^+$ for three 
flavors of neutrino as a function of $s$ the squared center of mass 
energy. The threshold is at $s = (m_W + m_l)^2$.}
\label{fig:nugam}
\end{figure}

\figref{fig:nugam} shows the cross-sections for 
$\nu \gamma \rightarrow l W^+$ for the three different neutrino flavors. 
Near threshold, the lepton propagator in \figrefab{fig:fd}{a} leads to a 
large logarithm which enhances the cross-section for $\nu_e$ over that 
for $\nu_\mu$ and $\nu_\tau$. Setting the lepton mass to zero everywhere 
but in the logarithm, the cross-section is fairly compact:
\ealab{eq:sig1a}{
\sigma_{\nu \gamma \rightarrow l W^+}& = & 
\sqrt{2} \alpha G_F \left[ 
2(1-\frac{1}{y})(1+\frac{2}{y^2}-\frac{1}{y^2}\log y) \right. + \nonumber \\  
& & \left. \frac{1}{y}(1 - \frac{2}{y}+\frac{2}{y^2})\log\frac{m_W^2 
(y-1)^2}{m_e^2 y} \right] \,,
}
where $y=s/m_W^2$ and $s=(p_\nu + p_\gamma)^2$, $G_F$ is Fermi's 
constant, and $\alpha$ is the fine structure constant which runs to 
$\sim 1/128$ near $m_W$.

\begin{figure}[ht]
\centering\leavevmode
\epsfxsize=\figwidth
\epsfbox{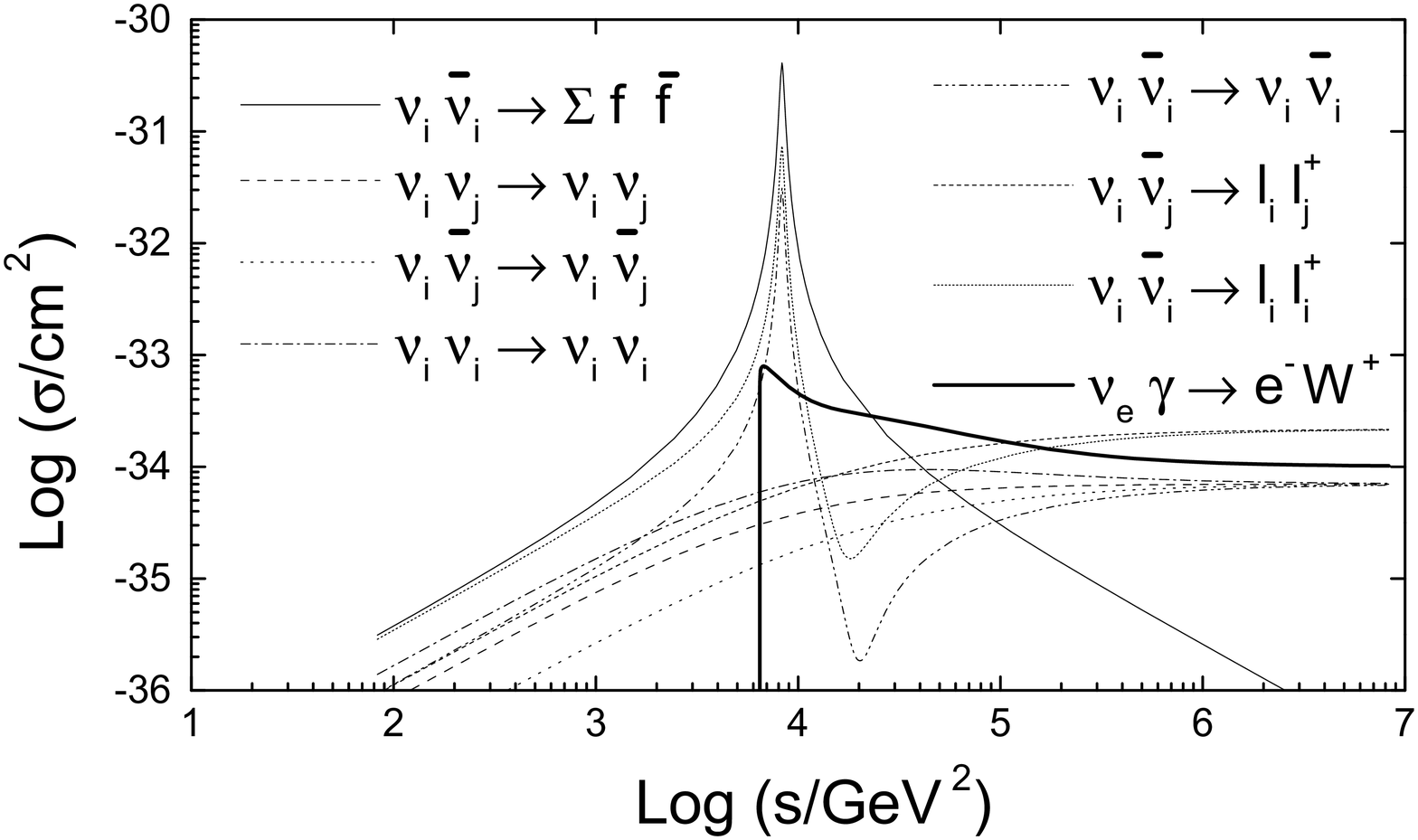}
\smallskip
\caption{Comparison of the $\nu_e\gamma$ cross-section to that for various $\nu\nu$ 
and $\nu\anti\nu$ processes as a function of $s$. The sum 
$\sum_j f_j \anti f_j$ does not include $f_j = \nu_i\,,l_i\,,t\,,W$, or 
$Z$.}
\label{fig:comp}
\end{figure}

It seems that the most interesting 
application of $\nu \gamma$ scattering is absorption of ultra high energy 
neutrinos off the microwave photon background. The potential importance of this 
process is illustrated in \figref{fig:comp}, where the $\nu_e \gamma$ 
cross-section is compared to relevant $\nu\nu$ and $\nu\anti\nu$ 
cross-sections at the same center of mass energies\CI{Roulet}. It is assumed that 
the neutrino mass is zero. The figure is  
dominated by processes involving intermediate $Z$ bosons at resonance, 
but at higher energies the $\nu \gamma$ cross-section is comparable or 
larger than that for the $\nu \nu$ reactions. 

\begin{figure}[ht]
\centering\leavevmode
\epsfxsize=\figwidth
\epsfbox{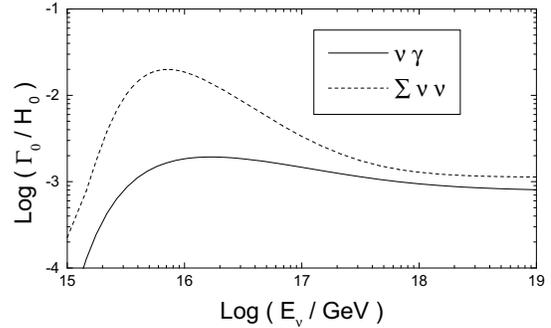}
\smallskip
\caption{Comparison of the absorption of high energy neutrinos from the 
cosmic background of photons and neutrinos to the cosmic expansion rate. 
The absorption by neutrinos includes a sum over the processes in 
Fig.~(3). The photon temperature is taken to be $T_\gamma = 2.74$ K, and
$T_\nu = (4/11)^{1/3} T_\gamma$. The 
universe is assumed to be matter dominated with an expansion rate of 
$H_0 = 50$ km/sec/mpc.}
\label{fig:abs}
\end{figure}

In a cosmological setting, the absorption rate is calculated by 
integrating the cross-section over the distribution of the target 
species. There are six flavors of neutrino and several processes to sum 
over. On the other hand photons have two spin degrees of freedom, and are 
more numerous than neutrinos by virtue of their higher temperature 
and boson statistics. \figref{fig:abs} 
shows the absorption rate of high energy neutrinos from the cosmic background of
photons or neutrinos. When $E_\nu \approx M_Z^2/T_\nu$ neutrino 
absorption is dominated by the $Z$ resonance, but at higher energies 
$\nu \gamma$ is important. At these higher energies the $\nu\nu$ 
processes mostly result in charged and neutral leptons, whereas the 
$\nu\gamma$ process produces $W^+$ bosons which mostly decay to quarks. 
Thus, not only is the amplitude of the absorption modified, but also the 
character of the cascade products. 

In the present epoch these processes are important only for neutrinos 
with energies $E_\nu > 10^{16}$ GeV, and even then only a small fraction 
of the beam is absorbed. Pushing back, neutrinos produced with energy 
$E_\nu (1+z) > 10^{16}$ GeV at redshifts $(1+z)>10$ would have been 
absorbed in their production epoch. A full cascade calculation must be 
done\CI{Yoshida}, evolving the ultra high energy neutrino distribution to 
low enough energies that they can propagate to the present unabsorbed. 
That cascade will be somewhat modified by the inclusion of $\nu\gamma$ 
reactions.


\mysection{$\nu N \rightarrow N \lowercase{l} W^+$.}
\label{sec:nuN}
In addition to reactions with real photons, it is also possible to 
convert $\nu \rightarrow l W^+$ in an external electromagnetic field.
The most obvious case to consider is the coulomb field of a nucleus, 
where both significant field strength and momentum transfer are 
possible. 

In the rest frame of the target nucleus, the cross-section per nucleon
can be expressed as a convolution over scattering of the neutrino with 
the virtual photons in the coulomb field.
\elab{eq:sig2}{
d\sigma_{\nu N \rightarrow N l W^+}  = d\sigma'
    \frac{I'}{I} \frac{Z^2 e^2 m_N^2 F_N^2(q^2)}{A q^4} \frac{d^3 q}{(2\pi)^3 2 m_N}\,,
}
where $d\sigma'$ is as in \eref{eq:sig1} except that the real photon 
is replaced by a virtual photon of momentum $q$ and polarization 
$j_N^\mu /m_N$. Here the electromagnetic current of the nucleus is
defined as $e Z j_N^\mu$.
In the rest frame of the nucleus, the matrix element used in 
$d\sigma'$ is $|{\cal M'}|^2 = 4 {\cal J}_{00}$, since in this frame
$p_N^\mu = m_N \delta^{\mu 0}$ and we use $q^\mu J_\mu = 0$. 
In \eref{eq:sig2} the quantity $I$ refers to the $\nu N$ system and 
$I'$ refers to the $\nu$-virtual $\gamma$ system, so that $I'/I = q z/m_N$, 
where $z$ is the direction cosine between the incident neutrino and $q$; $Z$ and 
$A$ are the charge and atomic number of the nucleus; and $F_N$ is the 
form factor of the nucleus normalized to $F_N(0)=1$

\begin{figure}[ht]
\centering\leavevmode
\epsfxsize=\figwidth
\epsfbox{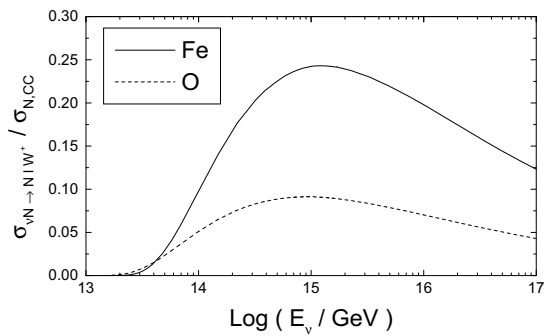}
\smallskip
\caption{Ratio of $\sigma_{\nu_e N \rightarrow N e W^+}$ to that for 
$\sigma_{\nu N, cc}$. The cross-sections are per nucleon. }
\label{fig:ratiocc}
\end{figure}

${\cal J}_{00}$ can then be expanded in powers of $q^2/m_W^2$, taking care to 
keep terms of order $E_\nu^2 q^2/m_W^4$  
until after the $d^3 q$ integration is done. In this expansion, $m_e$ may 
be safely set to zero as the logarithm associated with the intermediate 
lepton is cut off by $q^2$ which is generally larger than $m_e^2$. For 
$\nu_\mu$ and $\nu_\tau$ conversion, the lepton mass should be kept. 
The highest momentum components of the nuclear field establish the 
threshold for conversion. These have momenta of roughly $100$ MeV, so 
that $\nu N\rightarrow N l W^+$ has a threshold of $E_\nu \approx 10^{14}$ 
eV. This is an interesting range for current and proposed underwater/ice 
neutrino detectors. 
In \figref{fig:ratiocc} we show the ratio of the 
cross-section per nucleon for $\nu_e N \rightarrow N e W^+$, to that for 
charged current interactions\CI{GandhiQRS} for the cases where the 
nuclear target is oxygen and iron, as a function of neutrino energy.
The case of oxygen is 
interesting for neutrino detection rates in water (or ice) which are seen 
to increase by some 10\% at $E_\nu \approx 1$~Pev. The 
cross-sections on iron are increased by $20-25$\%, which will have an impact 
on studies of nucleon structure functions based on absorption of high 
energy cosmic neutrinos by the Earth. At higher energies, the charged 
current cross-section increases roughly as $E_\nu^{0.4}$\CI{GandhiQRS}
whereas the photon exchange process increases only logarithmically and 
becomes less important.


\mysection{$\anti\nu_{\lowercase{e}} \lowercase{e}^- \rightarrow \gamma 
W^-$.}
\label{sec:antinue}
Neutrino interactions in matter are usually dominated by scattering with 
nucleons. An exception is the case of $\anti\nu_e$: the $s$-channel 
reaction $\anti\nu_e e^- \rightarrow W^- \rightarrow f \anti f'$ is 
important near the $W$ resonance, although it decreases in importance at 
higher energy. Instead of the reaction with final state fermions $f\anti 
f'$, it is also possible to produce on-shell $W$'s accompanied by 
photons, $\anti\nu_e e^- \rightarrow \gamma W^-$, which is just the 
cross-channel of the $\nu_e \gamma \rightarrow e^- W^+$ reaction 
considered above. As long as one does not work too close to the resonance, 
the cross-section involves only the two diagrams related to those in 
\figref{fig:fd}. Dropping $m_e$ except in the logarithm, the result is
\ealab{eq:sig3}{
\sigma_{\anti\nu_e e^-}&&{}_{\rightarrow \gamma W^-} =  
 \frac{\sqrt{2} \alpha G_F}{3(y-1)y^2} \times \nonumber \\ 
& & \times  \left[ 3(y^2+1) \log(\frac{y m_W^2}{m_e^2}) 
-(5y^2+2y+5) \right] \,,
}
where $y=s/m_W^2$ and here $s=2 m_e E_\nu$.

\begin{figure}[ht]
\centering\leavevmode
\epsfxsize=\figwidth
\epsfbox{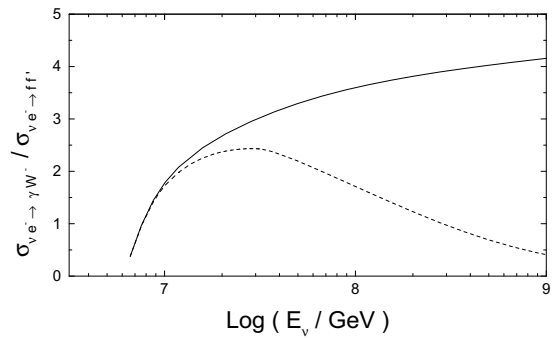}
\smallskip
\caption{Ratio of the cross-section for $\anti\nu_e e^- \rightarrow 
\gamma W^-$ to that for $\anti\nu_e e^- \rightarrow f \anti f'$. For the 
solid curve the sum over $f \anti f'$ includes only the $s$-channel to 
final states open in $W$ decay. The dashed curve includes the $b\anti t$ 
final state as well as the $t$-channel $Z$ exchange for elastic scattering.}
\label{fig:rationue}
\end{figure}

One might expect that with but a single channel and the smaller 
electromagnetic coupling that the $\gamma W^-$ reaction would be less 
important than $f \anti f'$ which proceeds to nine final states (12 above 
the top threshold). For very forward scattering, however, the $\anti\nu_e 
e^- \rightarrow \gamma W^-$ process involves the $t$-channel exchange of 
an almost on-shell electron, which leads to an enhancement by $\log 
s/m_e^2 \approx 25$. As a result the $\gamma W^-$ rate exceeds the 
$s$-channel rate to $f \anti f'$ summed over all species, as can be seen 
in \figref{fig:rationue}; i.e. the cross-section for $\anti\nu_e e^- 
\rightarrow \gamma f\anti f'$ exceeds that for $\anti\nu_e e^- 
\rightarrow f\anti f'$. 

At high energies, $t$-channel $Z$-boson exchange allows the elastic 
channel to dominate so the importance of $\gamma W^-$ decreases. For 
energies within a decay width of the resonance, a simple separation of 
the two processes is not possible - the photon is soft and so 
interference with initial and final state bremstrahlung emission must be 
considered\CI{WackerothH}. For energies outside the width of the 
resonance, the photon produced in $\anti\nu_e e^- \rightarrow \gamma W^-$ 
is hard and so will not interfere with bremstrahlung.


\mysection{Summary.}\label{sec:summary}
Neutrinos are generally considered to be weakly interacting particles, 
and thus neutrino-photon interactions are generally ignored, or confined 
to discussions of loop effects in scattering\CI{DicusR} or generating 
neutrino magnetic moments\CI{Akhmedov}. Here it is noted that for center 
of mass energies sufficient to produce real $W$ bosons that neutrinos and 
photons engage in $2 \rightarrow 2$ scattering  at tree level via lepton 
exchange and through W exchange and a $\gamma WW$ vertex. The resultant 
$\nu\gamma$ reactions are competitive with traditional neutrino reactions 
at high energies and in some cases may be dominant.

Three examples have been explored: a) $\nu\gamma \rightarrow l W^+$ is an 
important contribution to ultra high energy neutrino absorption in the 
early Universe, b) $\nu N \rightarrow N l W^+$ catalyzed by the nuclear 
coulomb field enhances $\nu N$ reaction rates by $10-20\%$ at neutrino 
energies of order $10^{15}$ eV, a range of interest to the next 
generation of neutrino telescopes, and c) the tree level process 
$\anti\nu_e e^- \rightarrow \gamma W^-$ is the dominant reaction in 
$\anti\nu_e e^-$ scattering for about a decade in $E_\nu$ just above the 
$W$ resonance. 

The current discussion has been confined to real $W$ production, but it 
should be apparent that neutrino-photon processes below threshold may 
also occur. Below threshold, however, the virtual $W$ must decay, which 
results in two suppressions: the final state will have three particles 
instead of two, which reduces the available phase space by about a factor 
of 100; and the decay vertex adds two powers of the weak coupling 
constant which is a further reduction by about a factor of 10. In a 
cosmological setting the process $\nu \gamma \rightarrow l f \anti f'$ 
would therefore be expected to be about a factor of 1000 smaller than 
corresponding $\nu \nu$ scattering at similar center of mass energies.

More interesting is the possiblity of $\nu N \rightarrow N l f \anti f'$ 
proceeding in the field of a nucleus. Even considering the $Z^2$ 
enhancement of the cross-section, the cross-section should be much 
reduced compared to the $\nu N$ charge exchange cross-section. However, 
for an experiment such as the detection of solar neutrinos at 
Kamiokande\CI{SUPERK} the relevant comparison is to $\nu e$ elastic 
scattering which is also suppressed relative to $\nu N$ scattering so it 
is unclear if `coulomb' scattering of neutrinos is unimportant. At 
slightly higher energies, even a modest contribution to $\nu N$ 
scattering in the few hundred MeV range could change interpretations of 
the atmospheric neutrino anomaly\CI{Gaisser}.

%
\mysection{Acknowledgments.}
I am grateful to S. Barr, S. Pittel and T. Stanev for helpful suggestions 
and advice. This work was partially supported by DOE grant 
DE-FG02-91ER40626. 

\end{document}